\documentclass[aps,prl,reprint,superscriptaddress,twocolumn,longbibliography]{revtex4-1}
\usepackage{amsfonts,amssymb,amscd,amsthm}
\usepackage{graphicx}
\usepackage{mathrsfs}
\usepackage[intlimits]{amsmath}
\usepackage[colorlinks, citecolor=red]{hyperref}
\usepackage{etoolbox}
\begin{document}
\title{Quantum simulation of the two-dimensional Weyl equation in a magnetic field}

\author{Y. Jiang}
\thanks{These authors contribute equally to this work}
\affiliation{Center for Quantum Information, Institute for Interdisciplinary Information Sciences, Tsinghua University, Beijing, 100084, PR China}

\author{M.-L. Cai}
\thanks{These authors contribute equally to this work}
\affiliation{Center for Quantum Information, Institute for Interdisciplinary Information Sciences, Tsinghua University, Beijing, 100084, PR China}
\affiliation{HYQ Co., Ltd., Beijing, 100176, P. R. China}

\author{Y.-K. Wu}
\thanks{These authors contribute equally to this work}
\affiliation{Center for Quantum Information, Institute for Interdisciplinary Information Sciences, Tsinghua University, Beijing, 100084, PR China}

\author{Q.-X. Mei}
\affiliation{Center for Quantum Information, Institute for Interdisciplinary Information Sciences, Tsinghua University, Beijing, 100084, PR China}

\author{W.-D. Zhao}
\affiliation{Center for Quantum Information, Institute for Interdisciplinary Information Sciences, Tsinghua University, Beijing, 100084, PR China}

\author{X.-Y. Chang}
\affiliation{Center for Quantum Information, Institute for Interdisciplinary Information Sciences, Tsinghua University, Beijing, 100084, PR China}

\author{L. Yao}
\affiliation{Center for Quantum Information, Institute for Interdisciplinary Information Sciences, Tsinghua University, Beijing, 100084, PR China}
\affiliation{HYQ Co., Ltd., Beijing, 100176, P. R. China}

\author{L. He}
\affiliation{Center for Quantum Information, Institute for Interdisciplinary Information Sciences, Tsinghua University, Beijing, 100084, PR China}

\author{Z.-C. Zhou}
\affiliation{Center for Quantum Information, Institute for Interdisciplinary Information Sciences, Tsinghua University, Beijing, 100084, PR China}

\author{L.-M. Duan}
\email{lmduan@mail.tsinghua.edu.cn}
\affiliation{Center for Quantum Information, Institute for Interdisciplinary Information Sciences, Tsinghua University, Beijing, 100084, PR China}

\begin{abstract}
Quantum simulation of 1D relativistic quantum mechanics has been achieved in well-controlled systems like trapped ions, but properties like spin dynamics and response to external magnetic fields that appear only in higher dimensions remain unexplored.
Here we simulate the dynamics of a 2D Weyl particle. We show the linear dispersion relation of the free particle and the discrete Landau levels in a magnetic field, and we explicitly measure the spatial and spin dynamics from which the conservation of helicity and properties of anti-particles can be verified.
Our work extends the application of an ion trap quantum simulator in particle physics with the additional spatial and spin degrees of freedom.
\end{abstract}

\maketitle

Relativistic quantum mechanics \cite{greiner2000relativistic,peskin2018introduction} combines two most important theories of modern physics: special relativity and quantum mechanics. It possesses negative-energy solutions which naturally lead to the prediction of antimatter, and it also explains the half-integer value of spins and their magnetic moments.
Weyl equation \cite{Weyl1929} is one of the simplest relativistic quantum mechanical equations. Its solution, a Weyl fermion, is a spin-$1/2$ particle with zero mass and a featured linear dispersion relation, and can be derived from the famous Dirac equation \cite{greiner2000relativistic,peskin2018introduction} in the massless limit. As one of the fundamental particles allowed by relativistic quantum mechanics \cite{greiner2000relativistic,peskin2018introduction}, the Weyl fermion was long believed to describe neutrinos, which are however extremely difficult to detect and are now known to have nonzero masses \cite{PhysRevLett.81.1562,PhysRevLett.87.071301}. This leaves many properties of the Weyl particles only to be analyzed theoretically, or through the idea of quantum simulation \cite{georgescu2014quantum,cirac2012goals} using other well-controlled quantum systems. Indeed, quantum simulation of relativistic quantum mechanical systems has been proposed \cite{Freedman2002,PhysRevLett.98.253005,PhysRevB.84.165115,doi:10.1126/science.1217069,PhysRevX.3.041018,Li2019} and performed \cite{Gerritsma2010,PhysRevLett.106.060503,Zhang2018,Kokail2019} in various physical systems like trapped ions \cite{RevModPhys.75.281,Blatt2012,doi:10.1063/1.5088164}.

To date, Weyl fermions have been realized in photonic crystals \cite{doi:10.1126/science.aaa9273} and in condensed matter systems \cite{doi:10.1126/science.aaa9297,PhysRevX.5.031013}, but in these systems only the spectral or the transport properties can be measured \cite{RevModPhys.90.015001}, while direct study of the Weyl particle dynamics is still lacking. On the other hand, massive Dirac particles have been simulated in ion trap \cite{Gerritsma2010,PhysRevLett.106.060503}, which can reduce to the massless Weyl particles by tuning the experimental parameters.
Nevertheless, to minimize the required degrees of freedom to be controlled, the experiments so far are restricted to dynamics in 1D, where interactions with external magnetic fields and evolution of spin states become trivial. In this work, we report the quantum simulation of a 2D Weyl fermion, which allows us to explore much richer spatial and spin dynamics.

Let us start from the 3D Dirac equation for a charged particle in a magnetic field \cite{greiner2000relativistic} (we have chosen the natural units by setting $\hbar=1$ and $c=1$ for simplicity)
\begin{equation}
i\frac{\partial \psi}{\partial t} = \hat{H} \psi = [\boldsymbol{\hat{\alpha}} \cdot (\boldsymbol{\hat{p}} - e \boldsymbol{\hat{A}}) + m \hat{\beta}] \psi,
\end{equation}
where $m$ is the mass of the particle, $e$ the electric charge, $\boldsymbol{\hat{A}}$ the vector potential of the magnetic field, and $\boldsymbol{\hat{p}}$ the momentum operator. $\hat{\alpha}_j$ ($j=x,y,z$) and $\hat{\beta}$ are Dirac matrices satisfying $\{\hat{\alpha}_i, \hat{\alpha}_j\}=2\hat{I}\delta_{ij}$, $\hat{\beta}^2=\hat{I}$ and $\{\hat{\alpha}_j,\hat{\beta}\}=0$. In 3D space, we thus need four matrices anti-commuting with each other, and therefore the Dirac matrices, as well as the Dirac spinor $\psi$, need to have a dimension of at least four \cite{greiner2000relativistic}.

\begin{figure}[!tbp]
   \includegraphics[width=\linewidth]{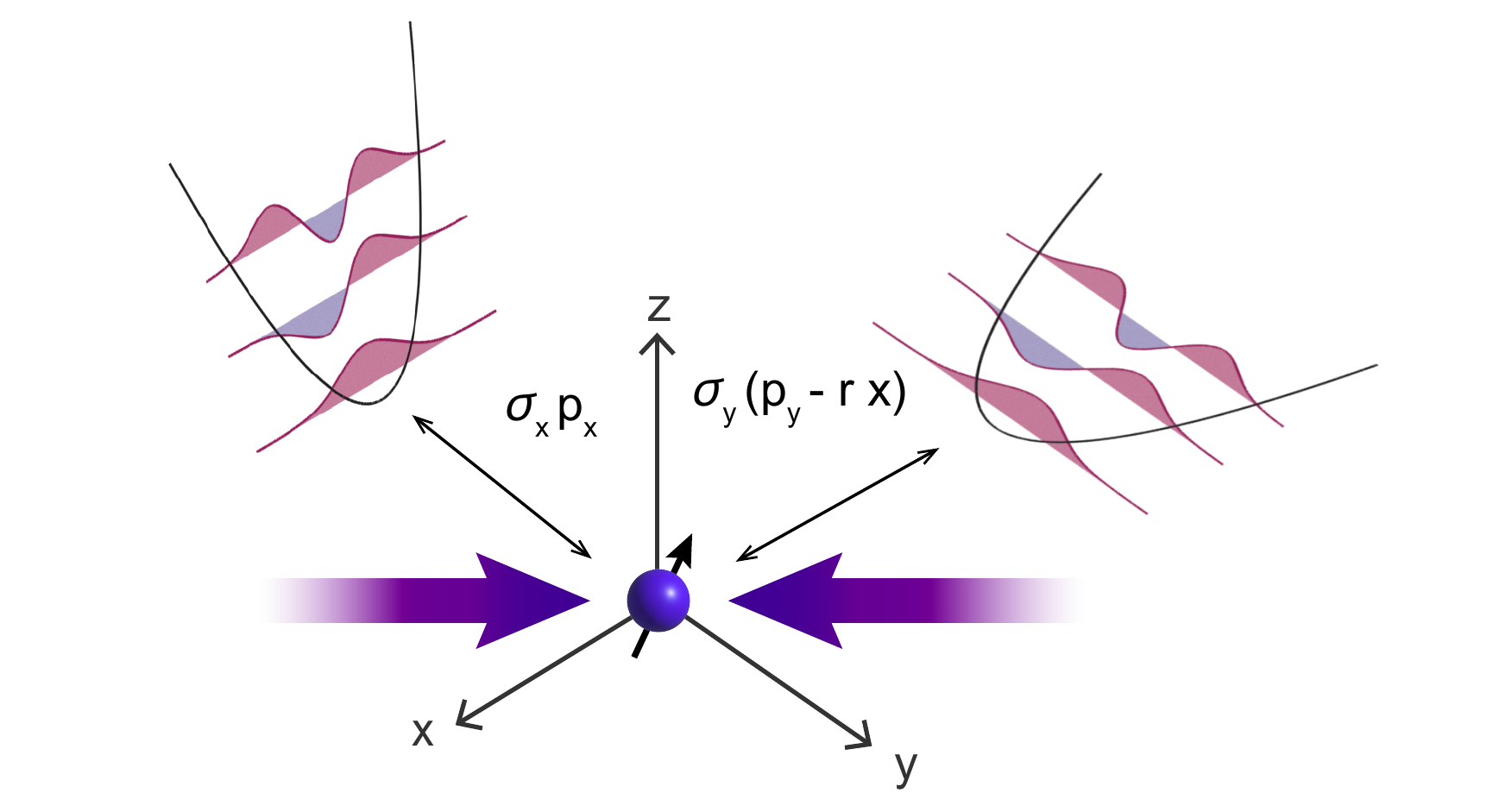}
   \caption{Schematic of simulating a 2D Weyl fermion using ion trap. We apply a pair of Raman laser beams on a trapped ion, with different frequency components resonant to the red and the blue sidebands of both the $x$ and $y$ oscillation modes, to couple the internal states of the ion to its spatial oscillation modes. By further tuning the phase of these frequency components in the Raman laser, we can choose to couple to $\hat{\sigma}_x$ or $\hat{\sigma}_y $ of the internal states, and to different quadratures $\hat{x}$ or $\hat{p}$ of each mode, which finally gives us the Hamiltonian in Eq.~(\ref{eq:H}).
   \label{fig:scheme}}
\end{figure}

\begin{figure*}[!tbp]
   \includegraphics[width=\linewidth]{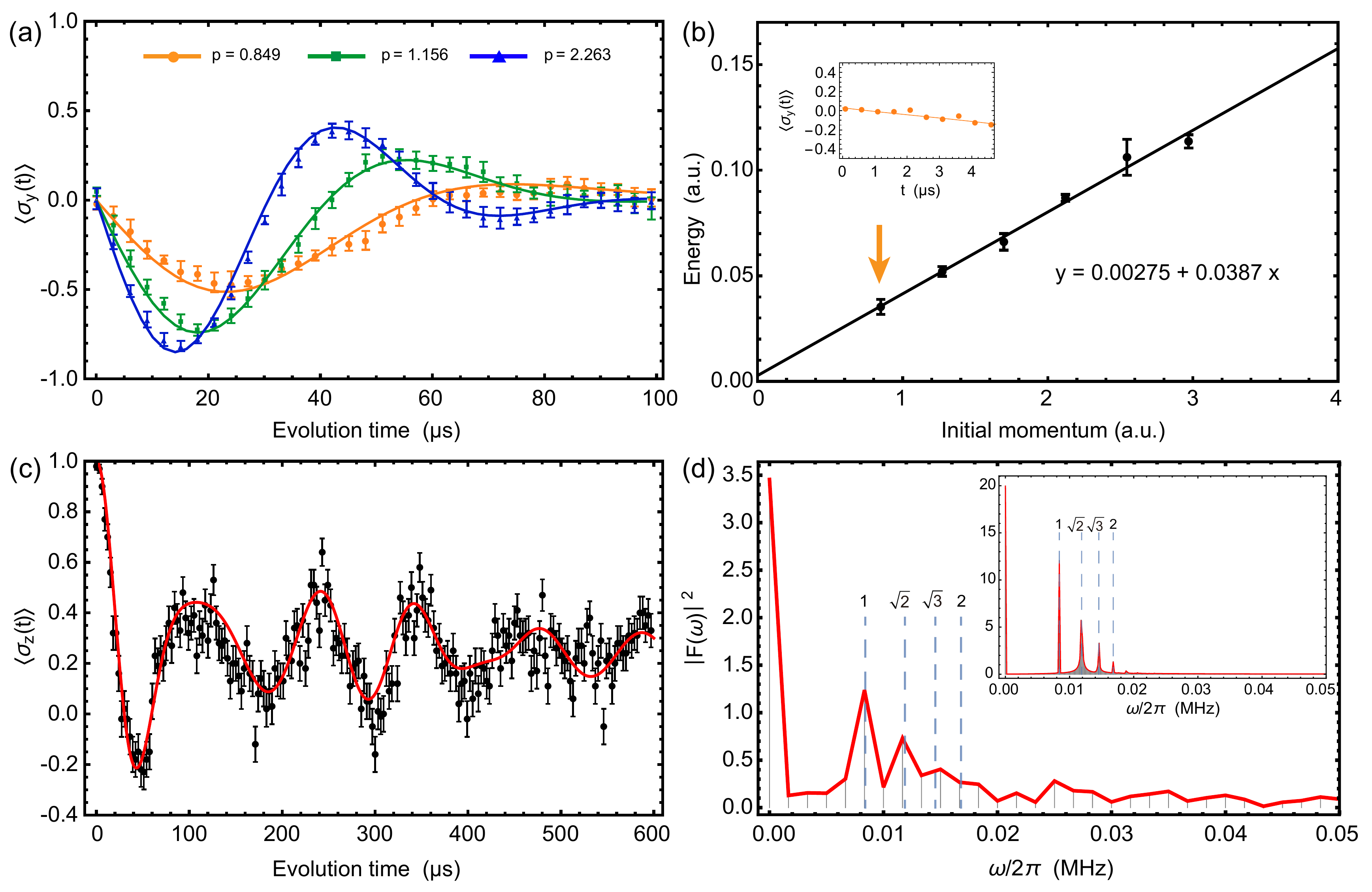}
   \caption{Linear dispersion relation of a free Weyl particle and its spectrum in a magnetic field. (a) Spin dynamics $\langle\hat{\sigma}_y(t)\rangle$ from an initial state $|+z\rangle|\alpha_x=ip/\sqrt{2}\rangle|\alpha_y=0\rangle$ with $B=0$ in Eq.~(\ref{eq:H}). Each data point consists of 500 experimental shots and the error bars are estimated from the standard deviation of 10 repetitions. The solid curves are theoretical predictions under the same parameters. (b) From the slope of the early-time spin dynamics (inset), we can extract the average energy $E(p)$ for a free Weyl particle with momentum $p$. A linear dispersion relation is observed as the black fitting line. The error bars are standard deviations of 5 repetitions. (c) Spin dynamics $\langle\hat{\sigma}_z(t)\rangle$ in a magnetic field with $eB=1$ from an initial state $|+z\rangle|\alpha_x=i\rangle|\alpha_y=0\rangle$. Due to the larger number of data points and the longer evolution time, here we only repeat each point by 200 times and the error bars are one standard deviation of the average value. The red curve is the theoretical prediction with the motional decoherence included. (d) The energy spectrum of a Weyl fermion in a magnetic field can be obtained by a Fourier transform of the spin dynamics. The first three peaks at $2E_0=0$, $2E_1=2\pi\times 8.3\,$kHz and $2E_2=2\pi\times 11.7\,$kHz agree well with the theoretical $E_n\propto \sqrt{n}$ scaling, while higher peaks are more difficult to distinguish due to the frequency resolution limited by the total evolution time of $600\,\mu$s. Inset is the ideal result for an evolution time of $5\,$ms without decoherence.
   \label{fig:spectrum}}
\end{figure*}

Now if we go to the massless limit $m\to 0$, the matrix $\hat{\beta}$ is removed and hence we only require three anti-commuting matrices. In this situation we can simply set $\hat{\alpha}_j=\hat{\sigma}_j$ ($j=x,y,z$) where $\hat{\sigma}_j$'s are the Pauli operators. This gives us the 3D Weyl equation \cite{greiner2000relativistic} in a magnetic field. We can further define a spin operator $\boldsymbol{\hat{S}}=\boldsymbol{\hat{\sigma}}/2$ such that the total angular momentum $\boldsymbol{\hat{J}}=\boldsymbol{\hat{L}}+\boldsymbol{\hat{S}}$ is conserved ($[\boldsymbol{\hat{J}}, \hat{H}]=0$) when the magnetic field is zero ($\boldsymbol{\hat{A}}=0$), where $\hat{L}_i=\epsilon_{ijk}\hat{x}_j \hat{p}_k$ is the orbital angular momentum and $\epsilon_{ijk}$ the Levi-Civita symbol \cite{greiner2000relativistic}.

Finally, for a minimal model to demonstrate nontrivial spatial and spin dynamics in a magnetic field, we consider a uniform field along the $z$ axis such that $\hat{A}_y=B\hat{x}$ and $\hat{A}_x=\hat{A}_z=0$. The momentum in the $z$ direction is conserved so that we can restrict to the 2D case, which gives us the Hamiltonian to be simulated
\begin{equation}
\label{eq:H}
\hat{H} = \hat{\sigma}_x \hat{p}_x + \hat{\sigma}_y (\hat{p}_y-eB\hat{x}).
\end{equation}

\begin{figure*}[!tbp]
   \includegraphics[width=\linewidth]{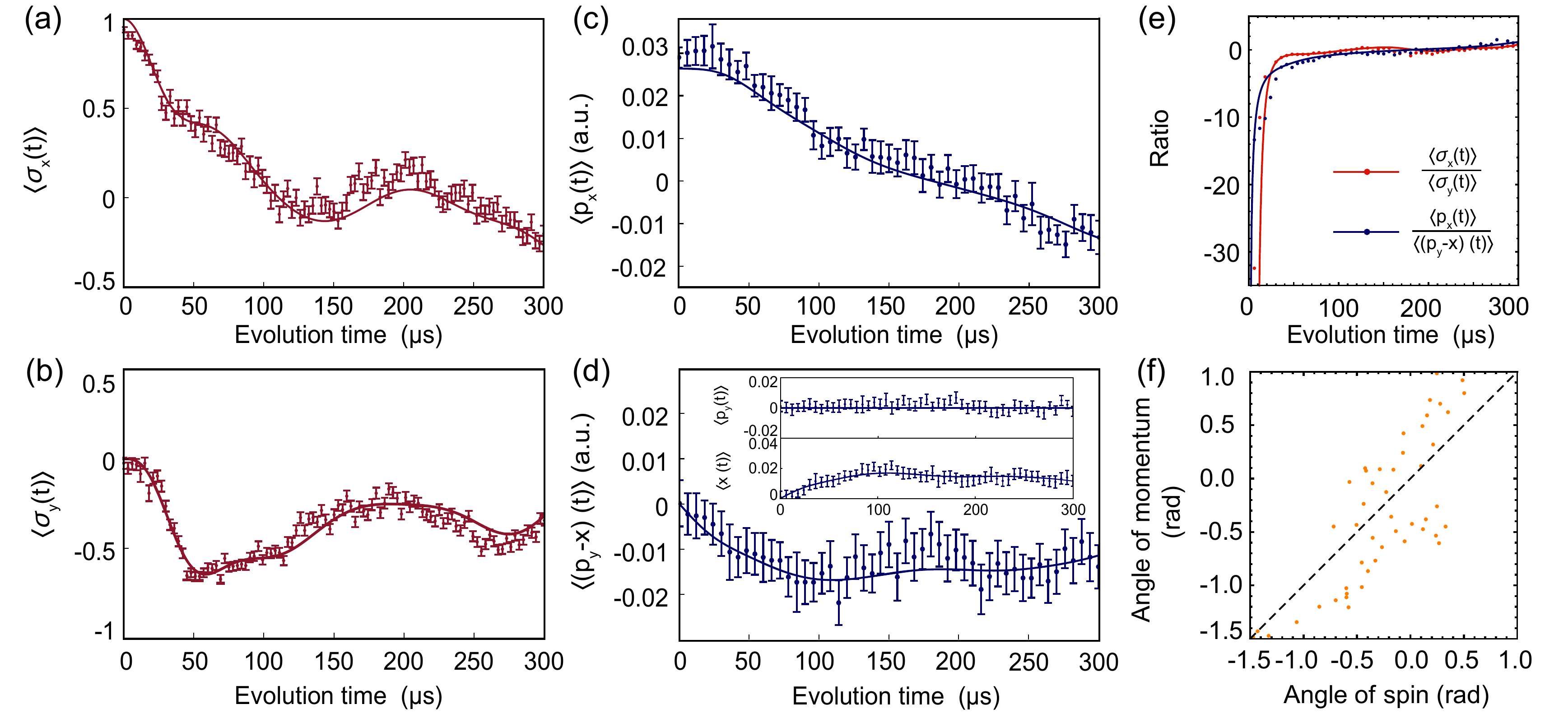}
   \caption{Spin and spatial dynamics and conservation of helicity. (a), (b) Spin dynamics in a magnetic field $eB=1$ from the initial state $|+x\rangle|\alpha_x=i\rangle|\alpha_y=0\rangle$. Each point is measured by 500 times and the error bar is estimated as one standard deviation of the average. Solid curves are theoretical results under the same parameters. (c), (d) Evolution of the kinetic momentum $\boldsymbol{\hat{\pi}}=\boldsymbol{\hat{p}}-e\boldsymbol{\hat{A}}$ under the same condition. Each quadrature is measured by applying an additional spin-dependent force and fitting the early-time evolution (see Supplementary Materials \cite{supplementary}), with error bars estimated from one standard deviation of the fitting. The error bar of $\langle\hat{\pi}_y\rangle=\langle\hat{p}_y\rangle-\langle\hat{x}\rangle$ is further computed from those of $\langle\hat{p}_y\rangle$ and $\langle\hat{x}\rangle$ as shown in the inset. (e) We plot the ratio between the $x$ and $y$ components of the spin (red) and the kinetic momentum (blue). The two values are close to each other and follow the same tendency when the ratios by themselves change orders of magnitudes, although small difference exists. Solid curves are the ideal theoretical results. (f) We further compute the azimuthal angles of the spin and the kinetic momentum in the $x$-$y$ plane. The data points distribute around the diagonal, which indicates that the spin and the kinetic momentum roughly align with each other during the time evolution in a magnetic field, namely the conservation of helicity.
   \label{fig:dynamics}}
\end{figure*}

Our experimental scheme is sketched in Fig.~\ref{fig:scheme}. A pair of counter-propagating $355\,$nm laser beams is shined on a trapped ${}^{171}\mathrm{Yb}^+$ ion to create a spin-dependent force \cite{PhysRevLett.112.190502,Rabi_model}. The Raman laser has an angle of $45^\circ$ to both the $x$ and the $y$ directions with trap frequencies $\omega_x=2\pi\times 2.35\,$MHz and $\omega_y=2\pi\times 1.98\,$MHz. By tuning the Raman laser resonant to the red (blue) sideband of the mode $j$ ($j=x,\,y$), we get the Hamiltonian $\hat{H}^{r(b)}_j(\Omega,\,\phi)=\Omega [\hat{\sigma}_{-(+)} \hat{a}_{j}^{\dag} e^{i \phi}+h.c.] / 2$ where $\hat{\sigma}_{+(-)}$ is the raising (lowering) operator of the qubit, $\hat{a}_j(\hat{a}_j^\dag)$ the annihilation (creation) operator of the corresponding mode, $\Omega$ the sideband Rabi frequency, and $\phi$ controlled by the phase of the Raman laser. Now we apply four frequency components, driving both sidebands of the two oscillation modes simultaneously as $\hat{H}=\hat{H}_x^r((1-r)\Omega,\,\pi/2)+\hat{H}_x^b((1+r)\Omega,\,\pi/2)+\hat{H}_y^r(\Omega,\,\pi)
+\hat{H}_y^b(\Omega,\,0) = (\Omega/\sqrt{2})[\hat{\sigma}_x \hat{p}_x + \hat{\sigma}_y(\hat{p}_y - r \hat{x})]$ where we define $\hat{x}=(\hat{a}_x+\hat{a}_x^\dag)/\sqrt{2}$ and $\hat{p}_x=i(\hat{a}_x^\dag-\hat{a}_x)/\sqrt{2}$ and similarly for the $y$ mode. This gives us the desired Hamiltonian with $r=eB$.

A characteristic property of the Weyl fermion is its linear dispersion relation in free space. With the above spin-dependent force, we can initialize the phonon states into $|\psi_m\rangle=|\alpha_x=i p_x/\sqrt{2}\rangle|\alpha_y=i p_y/\sqrt{2}\rangle$ with expectation values of the momentum as $\langle \hat{p}_x\rangle=p_x$ and $\langle \hat{p}_y\rangle=p_y$. Without squeezing, such a coherent state has a continuous distribution of the momentum, hence we also expect a continuous distribution of the energy of the Weyl particle and will focus on its expectation value. Note that, for a momentum on the $x$-$y$ plane with an angle $\tan\theta=p_y/p_x$, we have the positive-energy and negative-energy states when the spin is parallel or anti-parallel to it, namely the two eigenstates of $\hat{\sigma}_\theta\equiv\hat{\sigma}_x\cos\theta+\hat{\sigma}_y\sin\theta$. Now if we initialize the momentum state $|\psi_m\rangle$ with spin $|+z\rangle$, evolve the system under the free Weyl fermion Hamiltonian, and measure the early-time dynamics for a spin perpendicular to $\hat{\sigma}_\theta$ as $\hat{\sigma}_\theta^\perp \equiv -\hat{\sigma}_x\sin\theta+\hat{\sigma}_y\cos\theta$, we get $\langle \hat{\sigma}_\theta^\perp(t)\rangle \approx -t [ (\langle +\sigma_\theta | \langle \psi_m |) \hat{H} (|+\sigma_\theta\rangle |\psi_m\rangle) - (\langle -\sigma_\theta | \langle \psi_m |) \hat{H} (|-\sigma_\theta\rangle |\psi_m\rangle) ] = -2 E(p) t$ (see Supplementary Materials \cite{supplementary}). For example, in Fig.~\ref{fig:spectrum}(a) we plot the spin dynamics $\langle\sigma_y(t)\rangle$ from the initial state $|+z\rangle|\alpha_x=i p/\sqrt{2}\rangle|\alpha_y=0\rangle$ under various values of $p$, from which we extract a linear dispersion relation $E(p)\propto p$ as shown in Fig.~\ref{fig:spectrum}(b).

With the existence of a magnetic field, the continuous spectrum collapses into discrete Landau levels \cite{Landau1930}. For the massless Weyl particle, these energy levels have a unique scaling $E_n^{\mathrm{Weyl}}= \sqrt{2neB}$ compared with the Dirac particle $E_n^{\mathrm{Dirac}}= \sqrt{m^2 + 2neB}$ \cite{Lamata_2011} which in the nonrelativistic limit gives us the well-known result of $E_n^{\mathrm{NR}}=neB/m$ \cite{Landau1930}. In Fig.~\ref{fig:spectrum}(c), we plot the spin dynamics $\langle\sigma_z(t)\rangle$ in a magnetic field $eB=1$, which is oscillating at the frequency difference between different energy levels. Through a Fourier transform, we can identify discrete peaks in the spectrum as shown in Fig.~\ref{fig:spectrum}(d). Theoretically, these peaks locate at $2E_n$ (inset, see Supplementary Materials \cite{supplementary}) where the factor of $2$ comes from the positive and negative energy states. In this experiment, the evolution time up to $600\,\mu$s (which should be much shorter than the decoherence time of the motional states) restricts our frequency resolution so that high energy levels cannot be distinguished, but the first three peaks at $n=0,\,1,\,2$ already show good agreement with the predicted $\sqrt{n}$ scaling.

Next we examine the spatial and spin dynamics of a Weyl particle in the magnetic field. As we can see from Fig.~\ref{fig:dynamics}(a)-(d), in general, the components of the spin or the momentum are not conserved (except for $p_y$ in the inset which is related to our gauge choice). However, as shown in Fig.~\ref{fig:dynamics}(e) and (f), the directions of the spin $\boldsymbol{S}$ and the kinetic momentum $\boldsymbol{\pi}\equiv \boldsymbol{p}-e\boldsymbol{A}$ stay roughly the same during the evolution. This is known as the conservation of helicity $\hat{h}\equiv \boldsymbol{\hat{\sigma}}\cdot \boldsymbol{\hat{\pi}} / |\boldsymbol{\hat{\pi}}|$ \cite{greiner2000relativistic} in a magnetic field where $|\boldsymbol{\hat{\pi}}|$ represents the magnitude of the kinetic momentum. Strictly speaking, this conservation is proved for the scattering problem where the incoming and the outgoing particles do not feel the magnetic field \cite{PhysRevD.16.1815}, and from the theoretical curves we do observe small fluctuation between the orientations of these two vectors even under the ideal evolution. On the other hand, what we simulate here is more like the Larmor procession of a spin in a uniform magnetic field. By showing the spin procession speed to be matched by the spatial rotation rate $\omega=eB/m$ of the particle, this naturally explains the Lande $g$-factor of 2 for the spin-1/2 particles, which had to be added into the theory by hand before the relativistic quantum mechanics \cite{greiner2000relativistic}.

\begin{figure}[!tbp]
   \includegraphics[width=\linewidth]{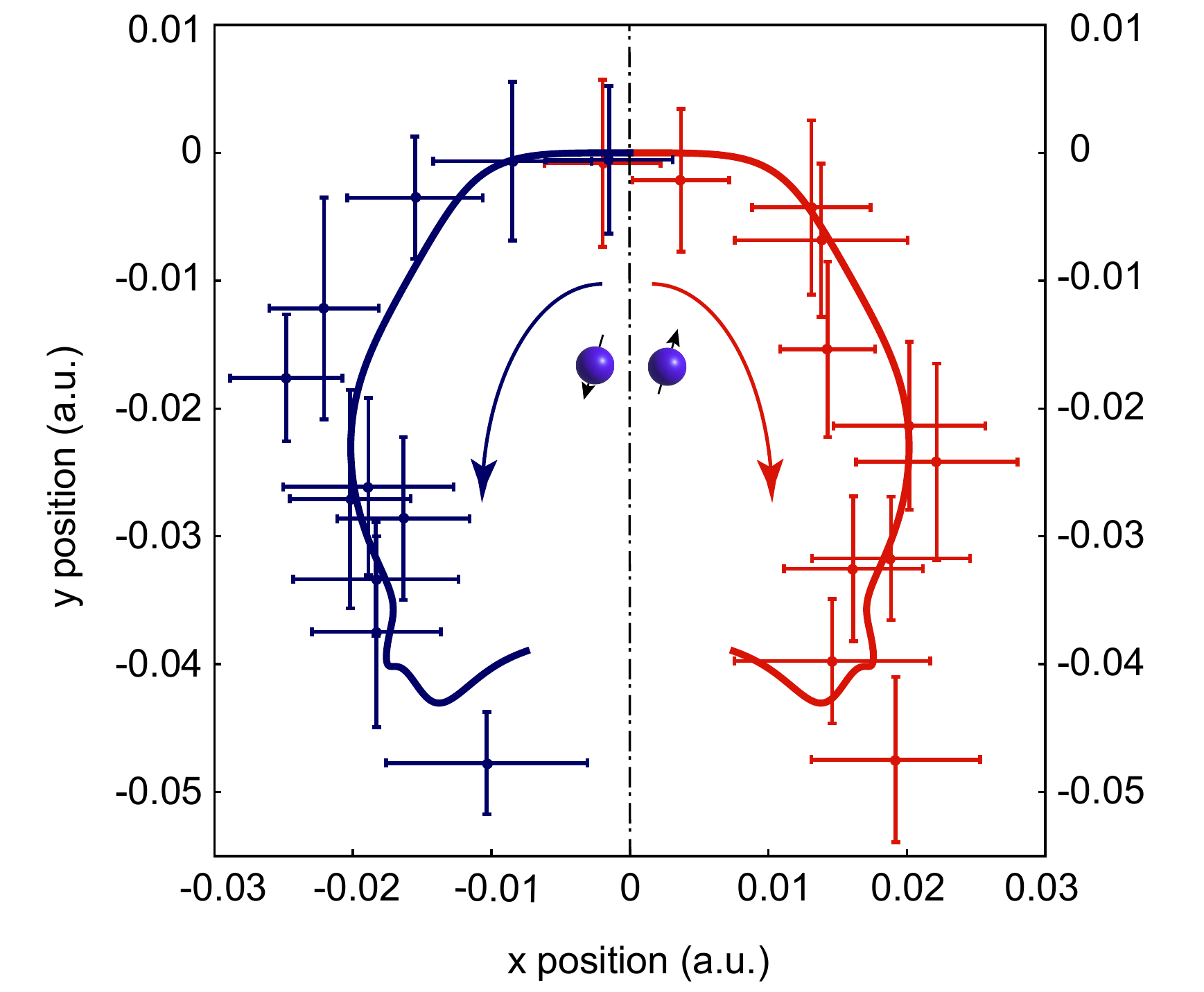}
   \caption{Trajectories of 2D Weyl particles with opposite helicities.  Using the method of Fig.~\ref{fig:dynamics}, we can plot the trajectory in a magnetic field $eB=1$ from the initial state $|+x\rangle|\alpha_x=i\rangle|\alpha_y=0\rangle$ (red) and $|-x\rangle|\alpha_x=i\rangle|\alpha_y=0\rangle$ (blue). These correspond to a Weyl particle with positive helicity and an anti-particle with negative helicity, respectively. Theoretically, the particles circles in real space with 2D Zitterbewegung as shown by the solid curves. Here our measurement precision cannot resolve the Zitterbewegung, but it can clearly be seen that the particle and the anti-particle have opposite charges as they rotate in opposite directions in a magnetic field.
   \label{fig:trajectory}}
\end{figure}

Following the same method, we can measure the spatial dynamics of a Weyl fermion with different initial spin states and we plot the 2D trajectories in Fig.~\ref{fig:trajectory}. For an initial state $|+x\rangle|\alpha_x=i\rangle|\alpha_y=0\rangle$ (red), namely a helicity of $+1$, the trajectory starts to the right and goes clockwise. On the other hand, for an initial state $|-x\rangle|\alpha_x=i\rangle|\alpha_y=0\rangle$ (blue) with a helicity of $-1$, the trajectory points to the opposite direction and goes counterclockwise. This is because the two situations correspond to a particle and its antiparticle, respectively, with opposite masses and charges. By initializing them with the same momentum, we thus get opposite initial velocities, and they bend into different directions due to the different signs in their charges.

To sum up, we have simulated a 2D Weyl particle with characteristic spectral properties using a single trapped ion and two of its spatial oscillation modes. Different from a neutrino, the simulated particle has a nonzero charge and demonstrates nontrivial spatial and spin dynamics in a magnetic field. Therefore our scheme provides a direct approach to study the dynamics of a class of elementary particles allowed by relativistic quantum mechanics but have not been discovered in nature. Theoretically, there will also be 2D Zitterbewegung on the trajectories (solid curves in Fig.~\ref{fig:trajectory}), whose amplitudes are however comparable to the error bars in our measurement results and therefore require future efforts to improve the experimental precision. By further setting the trap frequencies in the three spatial directions to be comparable and by pointing the Raman laser to be at a nonzero angle to all these axes, we can generalize this scheme to a 3D Weyl particle with even richer dynamics. Besides, our scheme of using spatial oscillation modes in higher dimensions can also be generalized to multi-ion cases such as the quantum simulation of the spin model \cite{monroe2021programmable} and the spin-phonon coupled system \cite{mei2021experimental}. Our work thus demonstrates the trapped ion system as a powerful quantum simulator, and significantly extends its application in particle physics by providing more spatial and spin degrees of freedom beyond 1D.

\begin{acknowledgments}
This work was supported by Tsinghua University Initiative Scientific Research Program, Beijing Academy of Quantum Information Sciences, and Frontier Science Center for Quantum Information of the Ministry of Education of China. Y.-K. W. acknowledges support from the start-up fund from Tsinghua University.
\end{acknowledgments}

%

\end{document}


\title{Supplementary Materials for \\
``Quantum Simulation of the Two-Dimensional Weyl Equation in a Magnetic Field''}

\author{Y. Jiang}
\thanks{These authors contribute equally to this work}
\affiliation{Center for Quantum Information, Institute for Interdisciplinary Information Sciences, Tsinghua University, Beijing, 100084, PR China}

\author{M.-L. Cai}
\thanks{These authors contribute equally to this work}
\affiliation{Center for Quantum Information, Institute for Interdisciplinary Information Sciences, Tsinghua University, Beijing, 100084, PR China}
\affiliation{HYQ Co., Ltd., Beijing, 100176, P. R. China}

\author{Y.-K. Wu}
\thanks{These authors contribute equally to this work}
\affiliation{Center for Quantum Information, Institute for Interdisciplinary Information Sciences, Tsinghua University, Beijing, 100084, PR China}

\author{Q.-X. Mei}
\affiliation{Center for Quantum Information, Institute for Interdisciplinary Information Sciences, Tsinghua University, Beijing, 100084, PR China}

\author{W.-D. Zhao}
\affiliation{Center for Quantum Information, Institute for Interdisciplinary Information Sciences, Tsinghua University, Beijing, 100084, PR China}

\author{X.-Y. Chang}
\affiliation{Center for Quantum Information, Institute for Interdisciplinary Information Sciences, Tsinghua University, Beijing, 100084, PR China}

\author{L. Yao}
\affiliation{Center for Quantum Information, Institute for Interdisciplinary Information Sciences, Tsinghua University, Beijing, 100084, PR China}
\affiliation{HYQ Co., Ltd., Beijing, 100176, P. R. China}

\author{L. He}
\affiliation{Center for Quantum Information, Institute for Interdisciplinary Information Sciences, Tsinghua University, Beijing, 100084, PR China}

\author{Z.-C. Zhou}
\affiliation{Center for Quantum Information, Institute for Interdisciplinary Information Sciences, Tsinghua University, Beijing, 100084, PR China}

\author{L.-M. Duan}
\email{lmduan@mail.tsinghua.edu.cn}
\affiliation{Center for Quantum Information, Institute for Interdisciplinary Information Sciences, Tsinghua University, Beijing, 100084, PR China}

\maketitle

\makeatletter
\renewcommand{\thefigure}{S\arabic{figure}}
\renewcommand{\thetable}{S\arabic{table}}
\renewcommand{\theequation}{S\arabic{equation}}
\makeatother

\section{Experimental setup}
We trap a single ${}^{171}\mathrm{Yb}^{+}$ ion in a linear Paul trap, whose two hyperfine states in the ground-state manifold ${}^{2}\mathrm{S}_{1/2}$ are chosen as the qubit states $\left|-z\right\rangle \equiv \left|F=0,m_F=0\right\rangle$ and $\left|+z\right\rangle \equiv \left|F=1,m_F=0\right\rangle$. We use $370\,$nm laser with suitable detuning and polarizatioin for Doppler cooling, optical pumping and qubit state detection. We use counter-propagating $355\,$nm pulsed-laser beams to generate spin-dependent forces through Raman transitions resonant to the blue and red sidebands of the oscillation modes, and we also use these Raman beams for sideband cooling to the phonon numbers of $\bar{n}_x < 0.05$ and $\bar{n}_y < 0.1$. Two acousto-optic modulators (AOMs) are used to fine-tune the frequency and the amplitude of each frequency components. More details about the daily operations of the trapped ion can be found in our previous work \cite{Rabi_model}.

\section{Quadrature measurements}
We follow the method of Ref.~\cite{Gerritsma2010} to measure the expectation values of different quadratures of the $x$ and $y$ modes. For example, to measure $\langle \hat{x} \rangle$, first we prepare the desired state to be measured under the evolution of the Weyl particle Hamiltonian. Then we reset the qubit state to $|-z\rangle$ using a $3\,\mu$s $370\,$nm optical pumping pulse with a negligible effect on the motional state \cite{Gerritsma2010}, and further rotate it to $|+x\rangle$. Next we apply a probe Hamiltonian $\hat{H}_p=(\Omega_p/\sqrt{2})\hat{\sigma}_y\hat{x}$, with a sideband Rabi frequency $\Omega_p$. Observe that $e^{-i\hat{H}_p\tau} \hat{\sigma}_z e^{i\hat{H}_p\tau} = \cos(\sqrt{2}\Omega_p \tau \hat{x}) \hat{\sigma}_z + \sin(\sqrt{2}\Omega_p \tau \hat{x}) \hat{\sigma}_x$. Therefore, by measuring $\langle\hat{\sigma}_z(\tau)\rangle$  for various duration $\tau$, we can extract $\langle\hat{x}\rangle$ from the slope in the short $\tau$ limit. To get more accurate results, we fit the early-time curve by a third-order polynomial and then get the slope at $\tau=0$ as the first-order coefficient. By switching the phase of the laser, we can similarly measure $\langle\hat{p}_x\rangle$, and by driving the sidebands of the $y$ mode, the quadratures $\hat{y}$ and $\hat{p}_y$ can be obtained.

In the experiment, we set $\Omega_p=\Omega$, that is, the probe has the same sideband Rabi rate as the simulated Weyl Hamiltonian (when making this measurement, the Weyl Hamiltonian is turned off after the desired evolution time). Specifically, in Figs.~2(a), 2(b) of the main text for the free Weyl fermion, we set $\Omega=2\pi\times 4.75\,$kHz. In Figs.~2(c), 2(d) and in Fig.~3 for dynamics in a magnetic field, we use $\Omega=2\pi\times 4.2\,$kHz. Then in Fig.~4, $\Omega=2\pi\times 5.0\,$kHz is used. The small difference in the Rabi frequency is due to the available laser power which is shared with the other setups. Nevertheless, so long as the Rabi frequency stays constant in each experiment, it will not affect the experimental result and we can always rescale $\Omega$ by a corresponding change in the unit of evolution time.

\section{Measuring average energy of a free Weyl particle}
For a two-mode coherent state $|\psi_m\rangle=|\alpha_x=i p_x/\sqrt{2}\rangle|\alpha_y=i p_y/\sqrt{2}\rangle$ on the $x$-$y$ plane with an angle $\tan\theta=p_y/p_x$, we define $\hat{\sigma}_\theta\equiv\hat{\sigma}_x\cos\theta+\hat{\sigma}_y\sin\theta$ and $\hat{\sigma}_\theta^\perp\equiv -\hat{\sigma}_x\sin\theta + \hat{\sigma}_y\cos\theta$. As mentioned in the main text, we initialize the spin state in $|+z\rangle$ and the momentum state in $|\psi_m\rangle$, turn on the free Weyl fermion Hamiltonian for time $t$, and measure the early-time dynamics for $\hat{\sigma}_\theta^\perp$. With the freedom of phases in the basis states, we can set $|+z\rangle=(|+\sigma_\theta\rangle+|-\sigma_\theta\rangle)/\sqrt{2}$ and $\hat{\sigma}_\theta^\perp |\pm\sigma_\theta\rangle = \mp i|\mp \sigma_\theta\rangle$. Then we get
\begin{align}
\langle \hat{\sigma}_\theta^\perp(t)\rangle =& (\langle+z|\langle\psi_m|) e^{i\hat{H}t} \hat{\sigma}_\theta^\perp e^{-i\hat{H}t} (|+z\rangle|\psi_m\rangle) \nonumber\\
\approx & \langle +z | \hat{\sigma}_\theta^\perp |+z\rangle + \frac{it}{2} (\langle +\sigma_\theta | + \langle -\sigma_\theta |) \langle \psi_m | (\hat{H} \hat{\sigma}_\theta^\perp - \hat{\sigma}_\theta^\perp \hat{H}) (|+\sigma_\theta\rangle+|-\sigma_\theta\rangle) |\psi_m\rangle \nonumber \\
=& -t \left[ (\langle +\sigma_\theta | \langle \psi_m |) \hat{H} (|+\sigma_\theta\rangle |\psi_m\rangle) - (\langle -\sigma_\theta | \langle \psi_m |) \hat{H} (|-\sigma_\theta\rangle |\psi_m\rangle) \right] \nonumber \\
= & -t (E_+ - E_-) \nonumber\\
= & -2 E(p) t.
\end{align}
Therefore we can get the expectation value of the energy from the slope of the early-time $\langle \hat{\sigma}_{\theta}^\perp(t) \rangle$.

\section{Discrete Landau levels by Fourier transform}
Here we give the Landau levels of a Weyl particle in a magnetic field and describe how they can be measured from the Fourier transform of the spin dynamics. Let us start from Eq.~(2) of the main text
\begin{equation}
\hat{H} = \hat{\sigma}_x \hat{p}_x + \hat{\sigma}_y (\hat{p}_y-eB\hat{x}).
\end{equation}

First we make a displacement along the $x$ direction by an amount of $\hat{p}_y/eB$ through the transform $\hat{U}=e^{i \hat{p}_x\hat{p}_y/eB}$ and get
\begin{equation}
\hat{U}\hat{H}\hat{U}^\dag = \hat{\sigma}_x \hat{p}_x - eB \hat{\sigma}_y \hat{x}.
\end{equation}

Next, we define $\hat{x}=(\hat{a}+\hat{a}^\dag)/\sqrt{2eB}$ and $\hat{p}_x=i\sqrt{eB/2}(\hat{a}^\dag-\hat{a})$, which satisfy the commutation relation $[\hat{x},\hat{p}_x]=i$. The Hamiltonian now becomes
\begin{equation}
\hat{U}\hat{H}\hat{U}^\dag = \sqrt{\frac{eB}{2}} \left[i \hat{\sigma}_x (\hat{a}^\dag-\hat{a}) - \hat{\sigma}_y (\hat{a}+\hat{a}^\dag)\right] = \sqrt{2eB}(i\hat{\sigma}_+ \hat{a}^\dag - i\hat{\sigma}_-\hat{a}).
\end{equation}

Note that the above transformations only influence the phonon states, while the spin state is untouched. Therefore it produces the same spin dynamics as the original Hamiltonian, as well as the same energy spectrum. The transformed Hamiltonian is just a blue sideband coupling with the well-known eigenstates $|E_0\rangle\equiv|+z\rangle|0\rangle$ and $|E_{n+1}^\pm\rangle\equiv(|-z\rangle|n\rangle \pm i|+z\rangle|n+1\rangle)/\sqrt{2}$ with energies $E_n^\pm=\pm\sqrt{2neB}$ ($n=0,\,1,\,2,\,\cdots$).
Now we can expand an arbitrary quantum state into the eigenstates as
\begin{equation}
\left|\psi\right\rangle =c_0|E_0\rangle + \sum_{n=1}^\infty \left(c_{n,+} \left|E_n^+\right\rangle + c_{n,-} \left|E_n^-\right\rangle\right),
\end{equation}
and its time evolution is given by
\begin{equation}
e^{-i\hat{H}t}\left| \psi\right\rangle = c_0|E_0\rangle + \sum_{n=1}^\infty\left(c_{n,+} e^{-iE_n^+ t} \left|E_n^+\right\rangle + c_{n,-} e^{-iE_n^- t} \left|E_n^-\right\rangle\right).
\end{equation}
Note that $\hat{\sigma}_z \left|E_n^\pm\right\rangle = -\left|E_n^\mp\right\rangle$ ($n\ge 1$) and $\hat{\sigma}_z |E_0\rangle=|E_0\rangle$, we thus have
\begin{equation}
\langle\psi| e^{i \hat{H} t}\hat{\sigma}_z e^{-i \hat{H} t} |\psi\rangle= |c_0|^2 - \sum_{n=1}^\infty \left[c_{n,+}^* c_{n,-} e^{i(E_n^+-E_n^-)t} + c.c.\right].
\end{equation}
Therefore the spin dynamics $\langle\hat{\sigma}_z(t)\rangle$ has frequency components of $E_n^+ - E_n^-=2\sqrt{2neB}$.

\section{Numerical simulations}
We use Qutip \cite{JOHANSSON20131234} to carry out the numerical simulations. In this experiment, motional dephasing is the dominant error source and is modelled by a Lindblad term $L[\sqrt{2/\tau_d}\hat{a}^\dag \hat{a}]\hat{\rho} \equiv (2/\tau_d) (\hat{a}^\dag \hat{a} \hat{\rho} \hat{a}^\dag \hat{a} - \hat{a}^\dag \hat{a} \hat{a}^\dag \hat{a}\hat{\rho} / 2 - \hat{\rho} \hat{a}^\dag \hat{a} \hat{a}^\dag \hat{a}/ 2)$,
where $\tau_d$ is the dephasing time. We adopt empirical values of $\tau_d=4\,$ms ($3.5\,$ms) for the $x$ ($y$) modes, respectively. In all the figures in the main text apart from Fig.~2(d) (inset) and Fig.~3(e), the theoretical curves are computed under the pre-calibrated parameters and the above motional dephasing time. The purpose of the theoretical curves in Fig.~2(d) (inset) and Fig.~3(e) is to demonstrate the ideal situation, and therefore for these curves the motional dephasing term is not added.

\begin{figure}[htbp]
   \includegraphics[width=\linewidth]{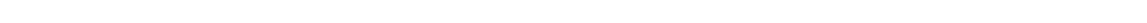}
   \label{fig:blank}
\end{figure}